\documentclass[preprint2]{aastex6}

\usepackage{booktabs}
\begin{document}



\title{Galactic Cosmic Ray Origins and OB Associations: Evidence from S\footnotesize{uper}\normalsize{TIGER} Observations of Elements $_{26}$F\MakeLowercase{e} through $_{40}$Z\MakeLowercase{r}}


\author{R. P. Murphy\altaffilmark{1}, M. Sasaki  \altaffilmark{2,6}, W. R. Binns\altaffilmark{1}, T. J. Brandt\altaffilmark{2}, T. Hams\altaffilmark{2,6}, M. H. Israel\altaffilmark{1}, A. W. Labrador\altaffilmark{3}, \\J. T. Link\altaffilmark{2}, R. A. Mewaldt\altaffilmark{3}, J. W. Mitchell\altaffilmark{2}, B. F. Rauch\altaffilmark{1}, K. Sakai\altaffilmark{2,6}, E. C. Stone\altaffilmark{3}, C. J. Waddington\altaffilmark{4}, \\N. E. Walsh\altaffilmark{1}, J. E. Ward\altaffilmark{1,7},} \and \author{M. E. Wiedenbeck\altaffilmark{5}}
\email{rmurphy@physics.wustl.edu}

\altaffiltext{1}{Department of Physics and McDonnell Center for the Space Sciences, Washington University, St. Louis, MO 63130, USA}
\altaffiltext{2}{NASA/Goddard Space Flight Center, Greenbelt, MD 20771, USA}
\altaffiltext{3}{California Institute of Technology, Pasadena, CA 91125, USA}
\altaffiltext{4}{University of Minnesota, Minneapolis, MN 55455, USA}
\altaffiltext{5}{Jet Propulsion Laboratory, California Institute of Technology, Pasadena, CA 91109, USA}
\altaffiltext{6}{Center for Research and Exploration in Space Science and Technology (CRESST), Greenbelt, MD 20771, USA}
\altaffiltext{7}{Now at Institut de Fisica d'Altes Energies (IFAE), Bellaterra (Barcelona), Spain}
\begin{abstract}
We report abundances of elements from $_{26}$Fe to $_{40}$Zr in the cosmic radiation measured by the SuperTIGER (Trans-Iron Galactic Element Recorder) instrument during 55 days of exposure on a long-duration balloon flight over Antarctica. These observations resolve elemental abundances in this charge range with single-element resolution and good statistics.
 These results support a model of cosmic-ray origin in which the source material consists of a mixture of 19$^{+11}_{-6}$\% material from massive stars and $\sim$81\% normal interstellar medium (ISM) material with solar system abundances. The results also show a preferential acceleration of refractory elements (found in interstellar dust grains) by a factor of $\sim$4 over volatile elements (found in interstellar gas) ordered by atomic mass (A). Both the refractory and volatile elements show a mass-dependent enhancement with similar slopes.

\end{abstract}




\section{Introduction} \label{sec:intro}
The SuperTIGER (Trans-Iron Galactic Element Recorder) instrument \citep{ST_InstrumentPaper}  was flown on a NASA long-duration balloon flight over Antarctica for 55 days in the 2012-2013 austral summer at altitudes from about 36.6 km to 39.6 km and a mean atmospheric overburden of 4.4 g cm$^{-2}$. The instrument measured the elemental abundances of Galactic cosmic-ray (GCR) nuclei with $10 \leq Z \leq 40$ above $\sim$700 GeV nucleon$^{-1}$ (at the top of the atmosphere).  

In this paper we present analysis of measurements of the elemental composition of ``ultra-heavy" GCRs with atomic number $26 \leq Z \leq 40$. These measurements are the first in which each element in the $30 \leq Z \leq 40$ charge range has been measured with single-element resolution and good statistics. 

In recent years, an explanation of GCR origins has emerged based on measurements from the 
Cosmic Ray Isotope Spectrometer (CRIS) \citep{StoneCRIS} onboard the NASA Advanced Composition Explorer (ACE) satellite \citep{StoneACE} and from the TIGER (Trans-Iron Galactic Element Recorder) balloon-borne instrument \citep{LinkThesis,RauchThesis,RauchPaper}. 

In this explanation, the GCR source material is thought to be a mixture of material from massive stars (supernova ejecta and stellar wind outflow), primarily within OB associations, and normal interstellar medium (ISM) material with solar system (SS) composition. These nuclei are then accelerated to cosmic-ray energies by supernova shocks. The possibility of OB association origin of GCR was first discussed by \citet{Reeves} and later developed by \citet{Hainebach}, \citet{CassePaul}, \citet*{Cesarsky}, and others. The previously measured composition of cosmic-ray isotopes and elements has been shown to be consistent with GCR origin in a source which is a mixture of $\sim$20\% material from massive star outflow and supernova ejecta, and $\sim$80\% material with SS abundances \citep{Higdon2003,BinnsNeon,RauchPaper,Binns_ACE}. For the remainder of this paper, the combined massive star wind outflow and supernova ejecta will be referred to as massive star material (MSM).

Observed GCR abundances show that elements that are found in interstellar dust grains (refractory elements) are preferentially accelerated compared to those that exist primarily as interstellar gases (volatile elements) \citep{MeyerDE,EllisonDM,MeyerEllison}. Elemental abundances measured by ACE at energies of hundreds of MeV nucleon$^{-1}$ \citep{Binns_ACE}, TIGER at GeV nucleon$^{-1}$ energies \citep{RauchPaper}, and CREAM at TeV nucleon$^{-1}$ energies \citep{CREAM_Ahn} show that this enhancement is mass dependent for both refractory and volatile elements, and that the ordering of these elements with atomic mass $A$ is greatly improved by comparing GCR source abundances with a mixture of normal ISM and MSM, rather than normal ISM alone (we note that \citet{EllisonDM} explain this mass-dependent trend only for the volatile elements). 

In this paper, we demonstrate that the elemental abundances  are consistent with a GCR source that consists of a best-fit mixture of 19$^{+11}_{-6}$\% MSM mixed with $\sim$81\% material with SS abundance, and an acceleration mechanism in which elements found in interstellar dust grains are preferentially accelerated over those found in interstellar gases.

Recent $\gamma$-ray observations of supernova remnants using the Fermi Large Area Telescope (LAT) \citep{FermiAcero} and ground-based imaging atmospheric Cherenkov telescope arrays such as HESS \citep{HESS_SuperBubble}, MAGIC \citep{MAGIC}, and VERITAS \citep{VERITAS_ICRC} have provided evidence of particle acceleration to very high energies. Moreover, observation of the $\pi^0$ turn on feature indicates that at least SNRs W44, IC443, and W51C are accelerating protons \citep{FermiAckermann2013,Jogler}, and all of these are believed to be the remnants of core-collapse supernovae. In addition, GeV and TeV emission has been observed from acceleration within the 30 Dor C superbubble in the Large Magellanic Cloud  \citep{HESS_SuperBubble}. Fermi has detected extended emission coinciding with a ``cocoon"-like morphology in the Cygnus superbubble \citep{Ackermann2011}, extending $\sim$50 parsecs from the Cygnus OB2 association. TeV emission from the Cygnus superbubble was also observed by ARGO-YBJ \citep{Bartoli}. These $\gamma$-ray observations lend further support to a model in which OB associations are a significant source of GCRs.

\section{Instrument Description} \label{sec:Instrument}
\begin{figure}[t]
\centering  
\includegraphics[width=0.5\textwidth]{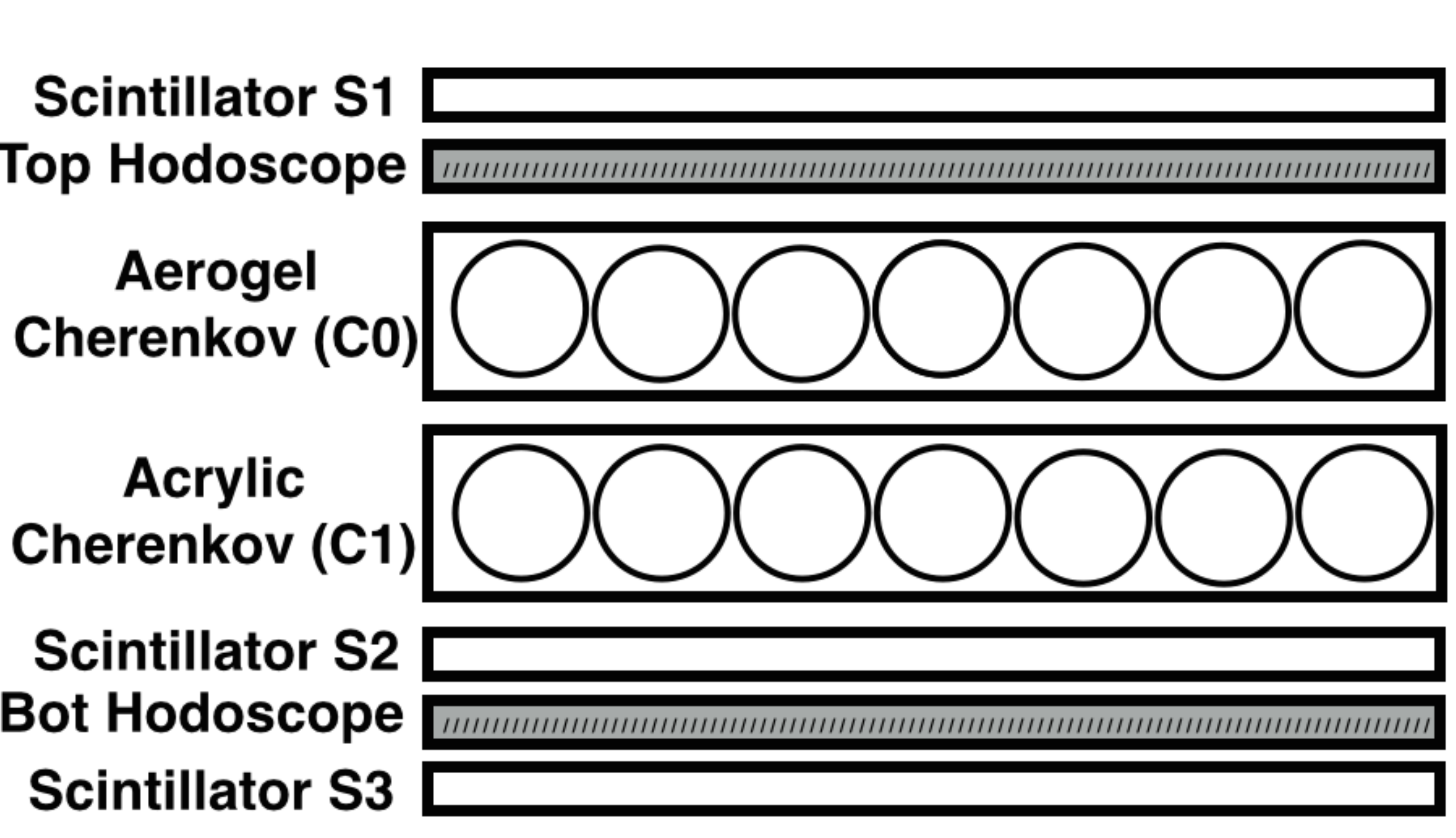}
\caption{Schematic side view of one SuperTIGER module.}
\label{ST_Instrument}
\end{figure}
The SuperTIGER instrument \citep{ST_InstrumentPaper,RPM_Thesis} consists of two nearly identical $\sim$1m $\times\ \sim$2m modules, each consisting of a suite of seven detectors. Figure \ref{ST_Instrument} shows a schematic side view of one module. Three scintillator detectors measure the differential energy loss $\frac{dE}{dx}$ within the instrument, which is a function of a particle's charge ($Z$) and velocity; two Cherenkov detectors, one with an aerogel radiator (C0), and one with an acrylic radiator (C1), which give signals that are different functions of $Z$ and velocity for particles above the Cherenkov threshold; and a scintillating fiber hodoscope (consisting of two separate x,y planes) which measures particle trajectory. For one SuperTIGER module, the Cherenkov radiators in the aerogel (C0) detector had an index of refraction $n=1.04$, while the other module had $n=1.04$ aerogel in half of the module and $n=1.025$ aerogel in the other half. The active area of each module measures approximately 1.16m$\times$2.4m, and the full geometry factor of both modules combined is $\sim$8.3 m$^2$ sr for particles whose trajectory zenith angle is less than 70 degrees. After accounting for losses due to nuclear interactions within the instrument, the ``effective" geometry factor is $ \sim $3.9 m$^2$ sr for $_{34}$Se.

\section{Data Analysis} \label{sec:Data}

\begin{figure}[t]
\centering  
\includegraphics[width=0.5\textwidth]{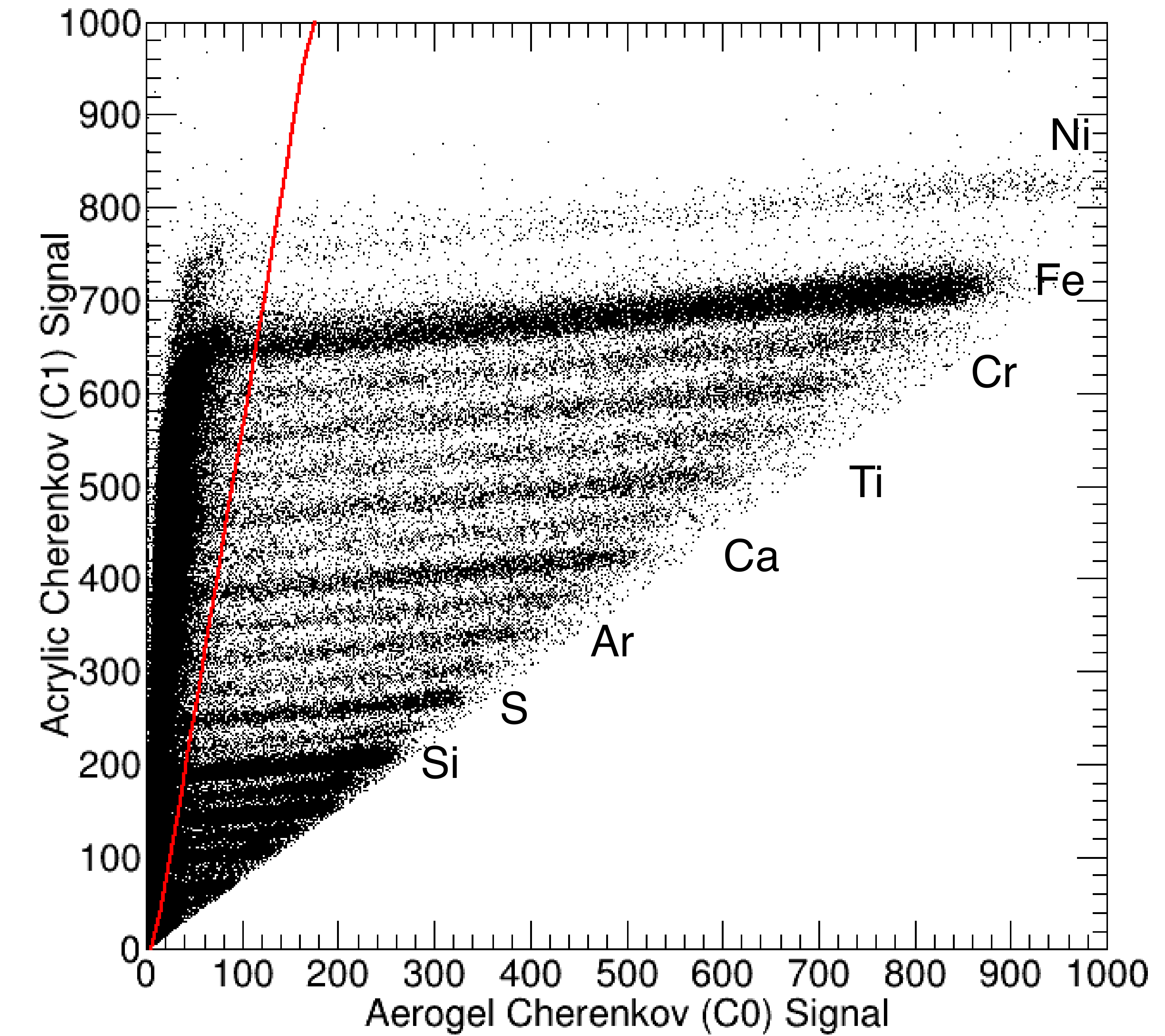}
\caption{Acrylic Cherenkov (C1) signal vs Aerogel Cherenkov (C0) signal cross plot for one day of data. Events to the left of the line have energy below or very close to the C0 threshold, and were analyzed using signals from the scintillator detectors and acrylic Cherenkov detector.}
\label{C1C0_crossplot}
\end{figure}
The charge ($Z$) of each cosmic-ray nucleus detected in the instrument was determined by one of two complementary techniques. At low energies (above the C1 threshold of $\sim$350 MeV nucleon$^{-1}$ but below the C0 threshold of $\sim$2.5 GeV nucleon$^{-1}$ or $\sim$3.3 GeV nucleon$^{-1}$, depending on the half-module the event went through), the charge was determined using a combination of signals from the top two scintillator detectors (S1 and S2) and the acrylic (C1) Cherenkov detector. At energies above the aerogel (C0) threshold, the charge was determined with a combination of the C1 and C0 detector signals. This technique was also used to analyze TIGER data from the 1997, 2001, and 2003 TIGER flights \citep{SposatoThesis,LinkThesis,RauchPaper}.

For each scintillator and Cherenkov detector, the signal was taken as the sum of the signals from all its photomultipliers. This sum was corrected for photomultiplier gain differences, temporal variations, and area nonuniformities using the $\sim5 \times 10^6$ $_{26}$Fe nuclei detected during flight to map the detector response. The particle trajectory from the scintillating fiber hodoscope was used to determine the particle position for the mapping correction. The trajectory angle $\theta$ with respect to the normal to the detector surface was used to correct for the sec$(\theta)$ dependence of signal on path length within the detector. Particles that underwent a charge-changing nuclear interaction within the instrument were identified and rejected by requiring agreement to within approximately one charge unit in the S1 scintillator, C1, and C0 Cherenkov detectors for events above the C0 threshold, and in the S1 and S2 scintillator detectors, and the C1 Cherenkov detector for events below the C0 threshold.
\begin{figure}[t]
\centering  
\includegraphics[width=0.5\textwidth]{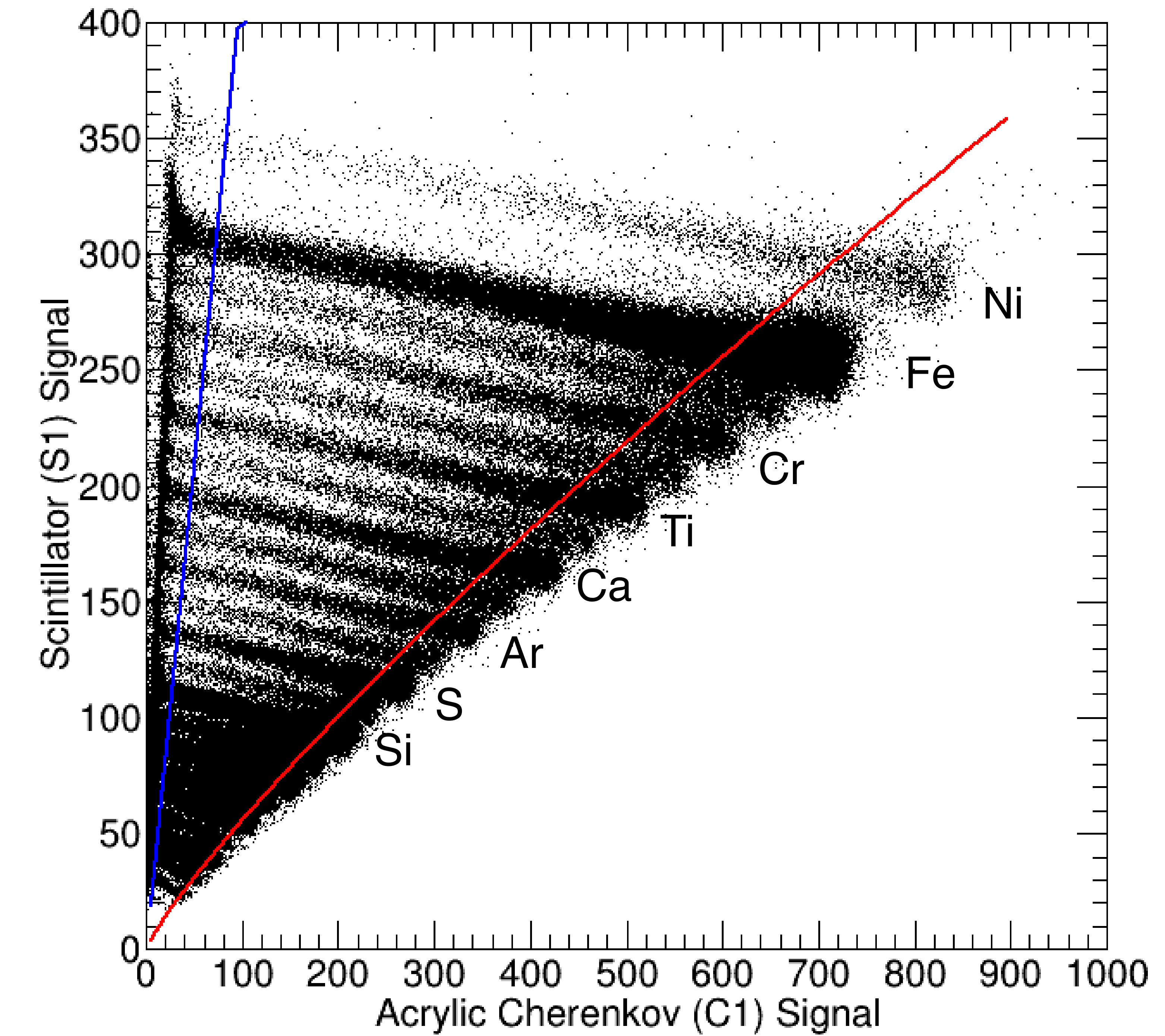}
\caption{Scintillator signal vs signal in acrylic Cherenkov detector cross plot for one day of data. Events to the right of the right (red) line have energy above the C0 threshold and were analyzed using the signal from the two Cherenkov detectors. Events to the left of the blue (left) line have energies near or below the C1 threshold and were discarded.}
\label{S1C1_crossplot}
\end{figure}
Figure \ref{C1C0_crossplot} is a cross plot of C1 vs C0 in which each point represents the signals from the two Cherenkov detectors for a single cosmic-ray nucleus (in this plot, only events and signals from the $n=1.04$ aerogels are shown). For particles with energies above the aerogel Cherenkov threshold, 
the combination of the two Cherenkov signals gives a well resolved charge assignment for events to the right of the red line. The points in Figure \ref{C1C0_crossplot} to the left of the line with low C0 signals represent events with energies near or below the aerogel Cherenkov threshold. Those events are analyzed using the signals from the scintillators and the acrylic Cherenkov detector.

Figure \ref{S1C1_crossplot} is a similar cross plot of signals from the top (S1) scintillator detector vs. the acrylic Cherenkov (C1) detector. The events to the right of the red (right) line have energies above the aerogel Cherenkov (C0) threshold, and were analyzed using signals from the two Cherenkov detectors as described previously. The events to the left of the left (blue) line are particles near or below the acrylic Cherenkov (C1) threshold of $\sim$ 350 MeV nucleon$^{-1}$, and are not included in the analysis. The events between the two lines were selected for analysis. For these lower-energy events, signals from the top two scintillator detectors (S1 and S2) and the acrylic Cherenkov (C1) detector were used to assign charge. The bottom (S3) scintillator detector was used to define a more restrictive data set for the development of these analysis techniques, but was not used in the final analysis. We fit curves of constant charge to each charge contour on the cross plot, and used those to fit an energy-independent form of the scintillator saturation model of \citet{VoltzModel} using the formalism given by \citet{Ahlen}. The resulting charge histogram was then renormalized so that charge peaks corresponded to their integer charges.

\begin{figure}[t]
\centering  
\includegraphics[width=0.5\textwidth]{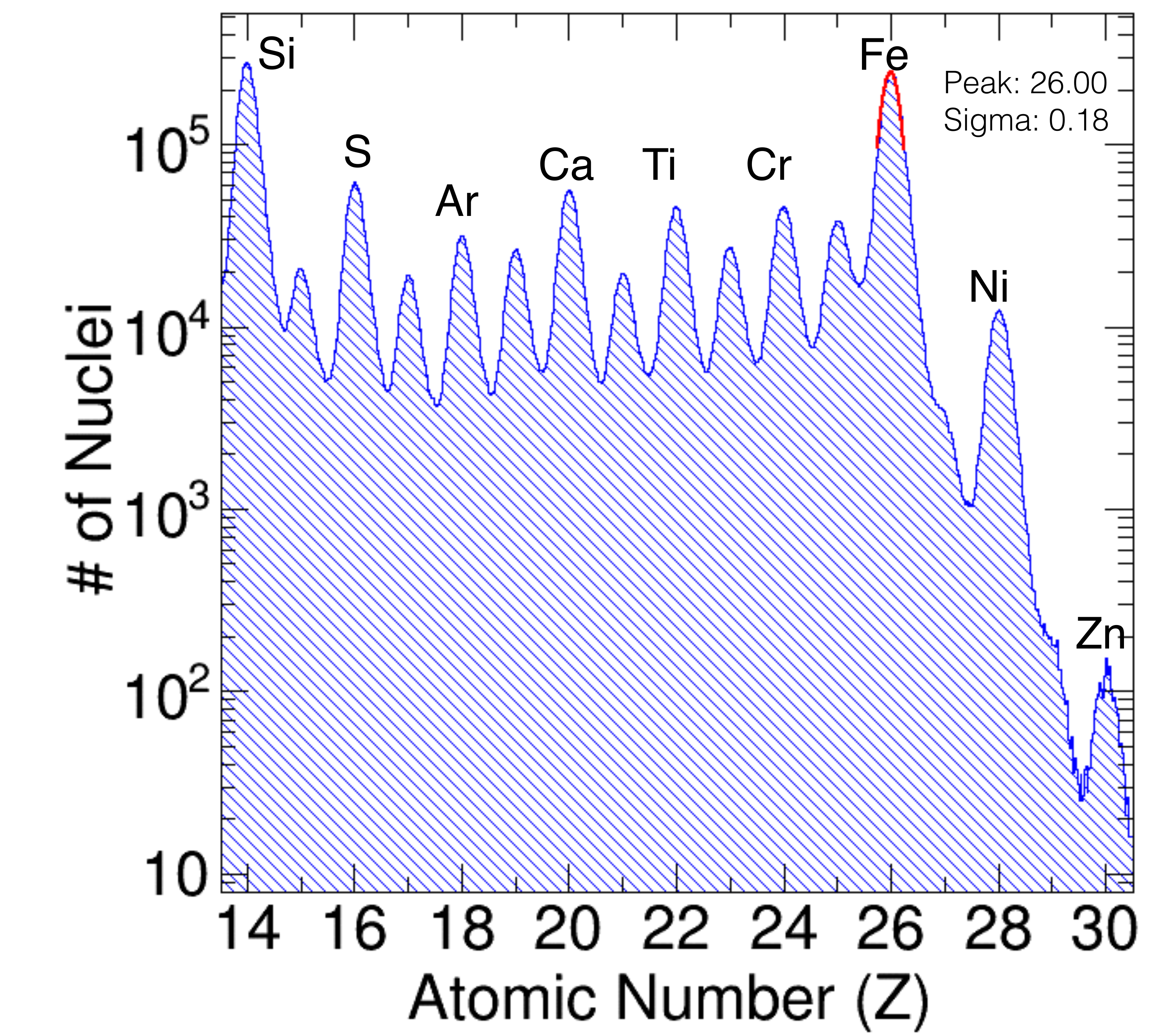}
\caption{Charge histogram showing SuperTIGER events analyzed with both high- and low-energy techniques from $_{14}$Si to $_{30}$Zn with 0.025 cu binning. The resolution at $_{26}$Fe is 0.18 charge units.}
\label{Charge_Hist_Full}
\end{figure}

\begin{figure}[t]
\centering  
\includegraphics[width=0.5\textwidth]{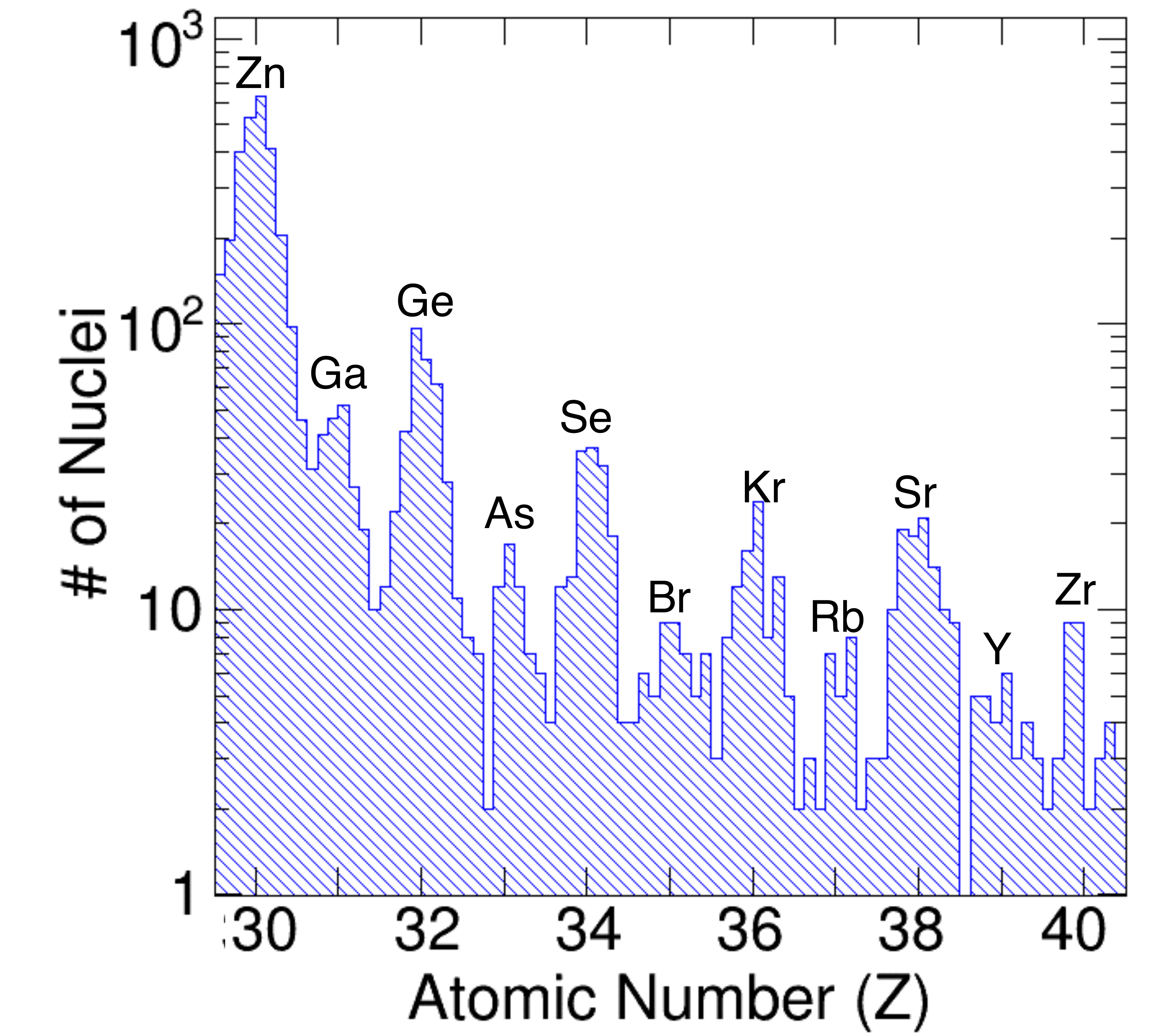}
\caption{Charge histogram showing SuperTIGER events analyzed with both high- and low-energy techniques from $_{30}$Zn to $_{40}$Zr with 0.125 cu binning.}
\label{Charge_Hist_UH}
\end{figure}

A charge histogram is shown in Figure \ref{Charge_Hist_Full}, which includes events analyzed with both techniques from the SuperTIGER flight. The 1$\sigma$ charge resolution at $_{26}$Fe is $0.18$ charge units (c.u.). Figure \ref{Charge_Hist_UH} uses a coarser binning of the data in the charge range $30 \leq Z \leq 40$, showing well defined, single-element peaks for every charge in that range. This histogram was fit with a multi-Gaussian function using a maximum likelihood method that was used to derive the measured abundances shown in column 3 of Table \ref{AbundanceTable}. 

\begin{table*}[t]
\caption{Cosmic Ray Element Abundances Relative to $_{26}$Fe = 10$^6$}
\label{AbundanceTable}
\begin{center}
\begin{tabular}{lc|r|rrr|rrr}
\hline
	&		&	\multicolumn{3}{r}{Observed in Instrument}    & & \multicolumn{3}{c}{Top-of-Atmosphere}		\\
Z	&	Element	&	Raw N	&	(Fe=10$^6$)	&	Upper Error	&	Lower Error	&	(Fe=10$^6$)	&Upper Error		&	Lower Error	\\
\hline
28	&	Ni	&	237391	&	50362	&	103	&	103	&	52880	&	3390	&	2620	\\
30	&	Zn	&	2623	&	556	&	11	&	11	&	619	&	44	&	35	\\
31	&	Ga	&	239	&	50.8	&	3.3	&	3.3	&	54.0	&	5.7	&	5.2	\\
32	&	Ge	&	354	&	75.1	&	4.0	&	4.0	&	86.8	&	8.2	&	6.9	\\
33	&	As	&	65	&	13.7	&	1.9	&	1.7	&	13.9	&	2.8	&	2.4	\\
34	&	Se	&	160	&	34.0	&	2.7	&	2.7	&	40.4	&	4.7	&	4.2	\\
35	&	Br	&	49	&	10.3	&	1.7	&	1.5	&	10.8	&	2.4	&	2.1	\\
36	&	Kr	&	91	&	19.4	&	2.2	&	2.0	&	24.1	&	3.7	&	3.1	\\
37	&	Rb	&	31	&	6.5	&	1.4	&	1.2	&	6.83	&	2.1	&	1.7	\\
38	&	Sr	&	105	&	22.3	&	2.2	&	2.2	&	29.8	&	3.9	&	3.6	\\
39	&	Y	&	30	&	6.4	&	1.4	&	1.2	&	7.80	&	2.1	&	1.7	\\
40	&	Zr	&	35	&	7.5	&	1.5	&	1.2	&	9.85	&	2.3	&	1.9	\\\hline
\end{tabular}
\end{center}
\par Column 3 lists the raw number of events observed in the SuperTIGER instrument; this corresponds to 4713661 $_{26}$Fe events in the same data set. Columns 4-6 show the abundances observed in the instrument and uncertainties relative to Fe=10$^6$. For the instrument abundances, uncertainties are statistical only. Statistical uncertainties for elements with raw number of events $N>100$  in the detector are $\sqrt{N}$ uncertainties; for elements with raw number of events $N<100$ we used the $\pm$1$\sigma$ errors given in Tables 1 and 2 of \citet{Gehrels}. These uncertainties have been renormalized with the abundances. Uncertainties in Top-of-Atmosphere abundances are total uncertainties, including propagated statistical uncertainties and the systematic uncertainties in interaction cross sections used for propagation calculations.
\end{table*}

\begin{figure}[t]
\centering  
\includegraphics[width=0.5\textwidth]{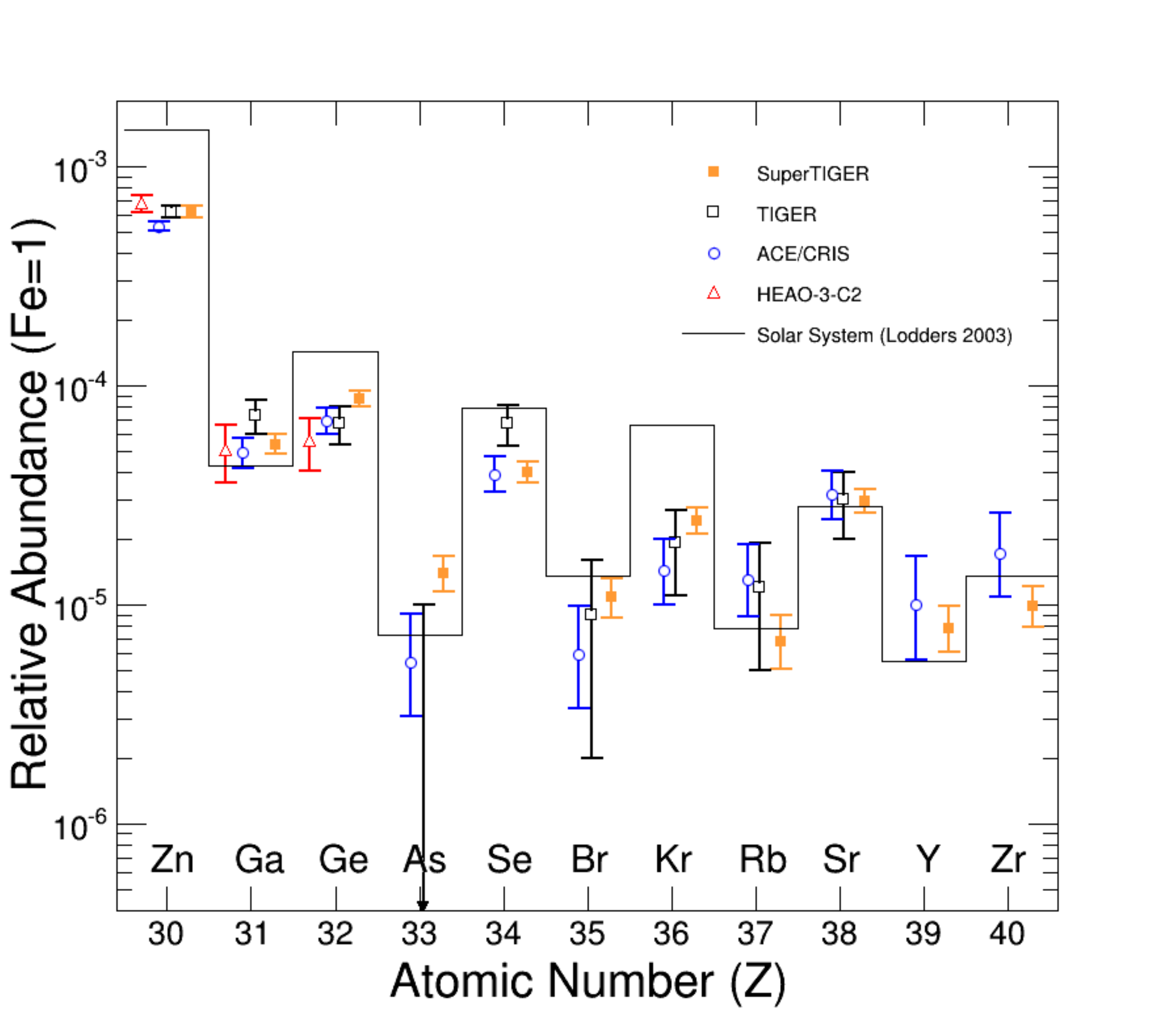}
\caption{Comparison of SuperTIGER Top-of-Atmosphere relative elemental abundances with abundances in space from ACE-CRIS \citep{Binns_ACE}, HEAO-3-C2 \citep{Byrnak}, and Top-of-Atmosphere abundances from TIGER \citep{RauchPaper}. Solar system elemental abundances \citep{Lodders2003} are also shown (solid lines). For the SuperTIGER points, combined statistical and systematic errors are shown.}
\label{TOA_Plot}
\end{figure}

We have derived abundances at the top of the atmosphere by correcting for the charge-dependent probability of particles undergoing nuclear interactions within the instrument and atmosphere, and the charge-dependent energy losses in the atmosphere and instrument. The correction for nuclear interactions within the instrument accounts for those particles identified and discarded during analysis. The atmospheric correction includes both the fraction of particles interacting ($\sim$36\% for $_{34}$Se) and secondary production. This was done using the same technique used for the TIGER data analysis, described in detail by \citet{RauchThesis}, using the total and partial charge-changing cross sections derived from accelerator data by \citet{Nilsen}. The derived Top-of-Atmosphere abundances are listed in columns 7-9 of Table \ref{AbundanceTable} and shown in Figure \ref{TOA_Plot}. That figure also shows GCR abundances measured in space by HEAO-3-C2 \citep{Byrnak},  ACE-CRIS \citep{Binns_ACE}, and Top-of-Atmosphere abundances measured by TIGER \citep{RauchPaper}. Also shown are SS elemental abundances \citep{Lodders2003}. For the SuperTIGER points, the combined statistical and systematic error bars are shown in solid orange. The systematic uncertainties for the SuperTIGER abundances were obtained by adjusting the total and partial charge changing cross-sections up and down by the uncertainty in those cross-sections (estimated by \citet{Nilsen}), and then comparing the results of the propagation calculation to the results obtained using unmodified cross sections. These systematic uncertainties are small compared to the statistical uncertainties. The ACE-CRIS error bars are purely statistical, while the TIGER error bars are calculated using Top-of-Instrument statistical uncertainties propagated through the atmosphere, and the HEAO-3-C2 error bars are statistical uncertainties, reported in \citet{Byrnak}.

\begin{table*}
\caption{Calculated Galactic Cosmic Ray Source Abundances Relative to $_{26}$Fe=10$^6$}
\label{GCRS_Table}
\begin{tabular}{ll|rrr|rrr}

\hline
	&		&	SuperTIGER Source	&		&		&	Combined Source	&		&		\\
Z	&Element		&	(Fe=10$^6$)	&	Upper Error	&	Lower Error	&	(Fe=10$^6$)	&	Upper Error	&Lower Error		\\
\hline
28	&	Ni	&	57600	&	3700	&	2860	&	57400	&	3330	&	2560	\\
30	&	Zn	&	658	&	50	&	40	&	655	&	45	&	36	\\
31	&	Ga	&	55.1	&	6.8	&	6.1	&	56.7	&	6.2	&	5.7	\\
32	&	Ge	&	86.0	&	9.1	&	7.8	&	82.5	&	8.4	&	7.2	\\
33	&	As	&	11.7	&	3.4	&	3.0	&	11.5	&	3.4	&	3.0	\\
34	&	Se	&	31.2	&	5.2	&	4.6	&	36.8	&	5.2	&	4.7	\\
35	&	Br	&	10.5	&	3.1	&	2.7	&	10.4	&	3.1	&	2.7	\\
36	&	Kr	&	17.1	&	4.3	&	3.7	&	16.3	&	4.0	&	3.4	\\
37	&	Rb	&	5.7	&	2.7	&	2.3	&	11.0	&	3.0	&	2.8	\\
38	&	Sr	&	31.7	&	4.8	&	4.4	&	31.7	&	4.5	&	4.1	\\
39	&	Y	&	10.3	&	2.9	&	2.4	&	10.2	&	2.9	&	2.4	\\
40	&	Zr	&	13.0	&	3.1	&	2.6	&	12.9	&	3.1	&	2.6	\\
\hline
\end{tabular}
Columns 3-5 show calculated SuperTIGER GCRS abundances and uncertainties. Columns 6-8 show the combined SuperTIGER and TIGER \citep{RauchPaper} GCRS abundances plotted in Figures \ref{GCRS_SS} and \ref{GCRS_MSO}.
\end{table*}

 The SuperTIGER data points generally agree with previous experiments, but have significantly smaller error bars. Of particular note are $_{31}$Ga and $_{32}$Ge. TIGER measurements indicated nearly equal abundances for these two elements, which was not expected in view of the high SS $_{32}$Ge/$_{31}$Ga ratio \citep{RauchPaper}.  While the SuperTIGER measurement of each of these elements is not in statistical disagreement with those of TIGER, SuperTIGER, with its much better statistics, shows that the abundances of these two elements are not equal; rather $_{32}$Ge/$_{31}$Ga is approximately 1.5, with a statistical difference in ratios of nearly 5$\sigma$. 

\section{Discussion} \label{sec:Discussion}
\begin{figure}[t]
\centering  
\includegraphics[width=0.5\textwidth]{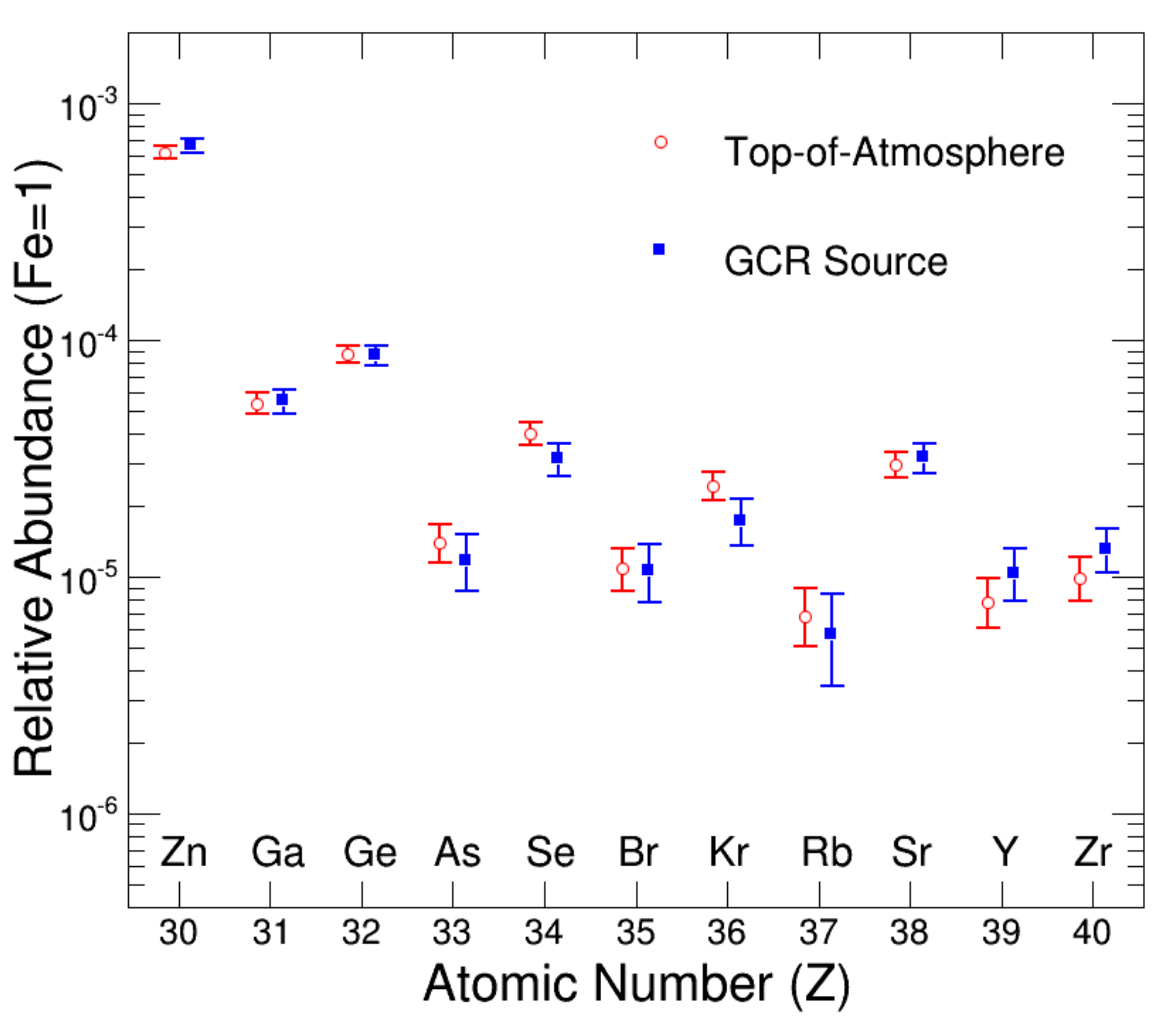}
\caption{Comparison of SuperTIGER Top-of-Atmosphere abundances and Galactic cosmic ray source (GCRS) abundances.}
\label{TOA_GCRS}
\end{figure}

Cosmic-ray source abundances were derived from the SuperTIGER Top-of-Atmosphere abundances using a leaky box propagation model \citep{Wiedenbeck_LeakyBox}, which uses total destruction cross sections (a modified form of those from \citet{Webber}) and partial cross sections from \citet{Silverberg}. The interstellar propagation results were used as input to a spherically symmetric modulation model based on \citet{Fisk}, with a modulation level $\phi = $ 543 MV and a typical Top-of-Atmosphere energy of $\sim$3.1 GeV nucleon$^{-1}$, to obtain modulated values for comparison with the abundances observed at Earth. This $\phi$ was inferred from spectra observed by ACE/CRIS during the SuperTIGER flight, which were measured using the method described in \citet{Wiedenbeck_Phi}.  The assumed cosmic-ray source abundances were adjusted to yield agreement with the data. The derived source abundances are shown in columns 3-5 of Table \ref{GCRS_Table}. The uncertainties reported in Table \ref{GCRS_Table} are the propagated Top-of-Atmosphere uncertainties. Figure \ref{TOA_GCRS} shows the Top-of-Atmosphere abundances reported in Table \ref{AbundanceTable} compared to the calculated source abundances.

\begin{figure}[t]
\centering  
\includegraphics[width=0.5\textwidth]{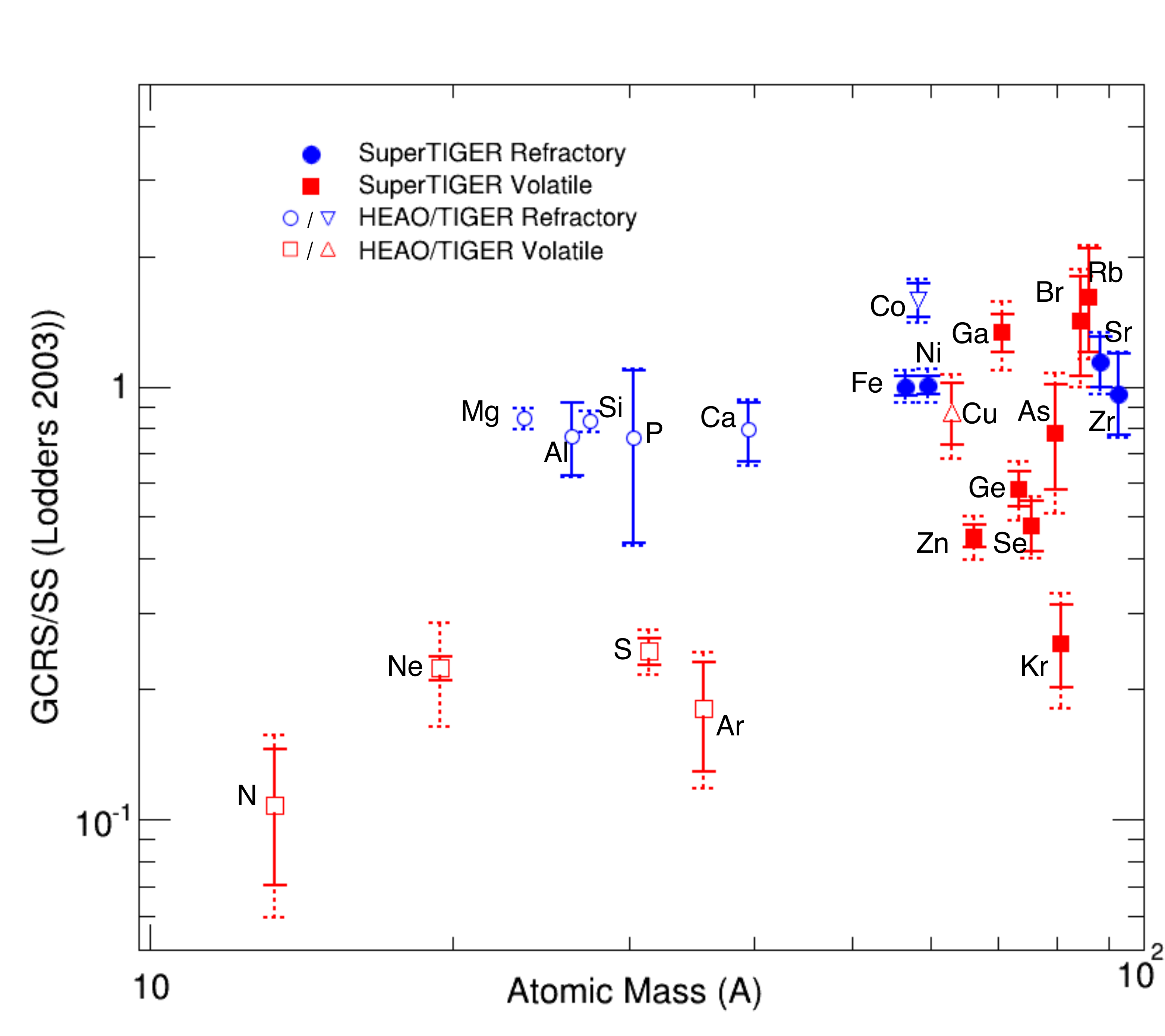}
\caption{Ratio of GCRS abundances to SS abundances \citep{Lodders2003} vs. atomic mass (A). Refractory elements are shown as blue circles; volatile elements are shown as red squares. Solid error bars show the uncertainty in the ratio due to uncertainty in the GCRS measurement; dashed error bars show the total uncertainty in the GCRS/SS ratio, including uncertainties in the SS abundances.}
\label{GCRS_SS}
\end{figure}
Figure \ref{GCRS_SS} is a plot of the ratio of Galactic cosmic-ray source (GCRS) elemental abundances to SS abundances from \citet{Lodders2003} as a function of atomic mass $A$. 
For elements with $Z<26$, the source abundances are those derived by \citet{Engelmann} from HEAO-3-C2 data. For $_{27}$Co and $_{29}$Cu, the abundances from \citet{RauchPaper} were used. For all other elements shown with $Z>26$, each point represents the source abundances calculated for SuperTIGER combined with TIGER abundances from \citet{RauchPaper}, weighted by the statistics recorded with each experiment. The refractory elements have equilibrium condensation temperatures \citep{Lodders2003} greater than $\sim$1200 K and the volatile elements have condensation temperatures lower than $\sim$1200 K. As noted by \citet{MeyerDE} and \citet{EllisonDM}, among others, the GCRS/SS ratio is generally higher for refractory elements than for volatile elements, especially at low $A$. However, at high $A$ the two groups merge and there is significant scatter, as noted by \citet{RauchPaper}.

\begin{figure}[t]
\centering  
\includegraphics[width=0.5\textwidth]{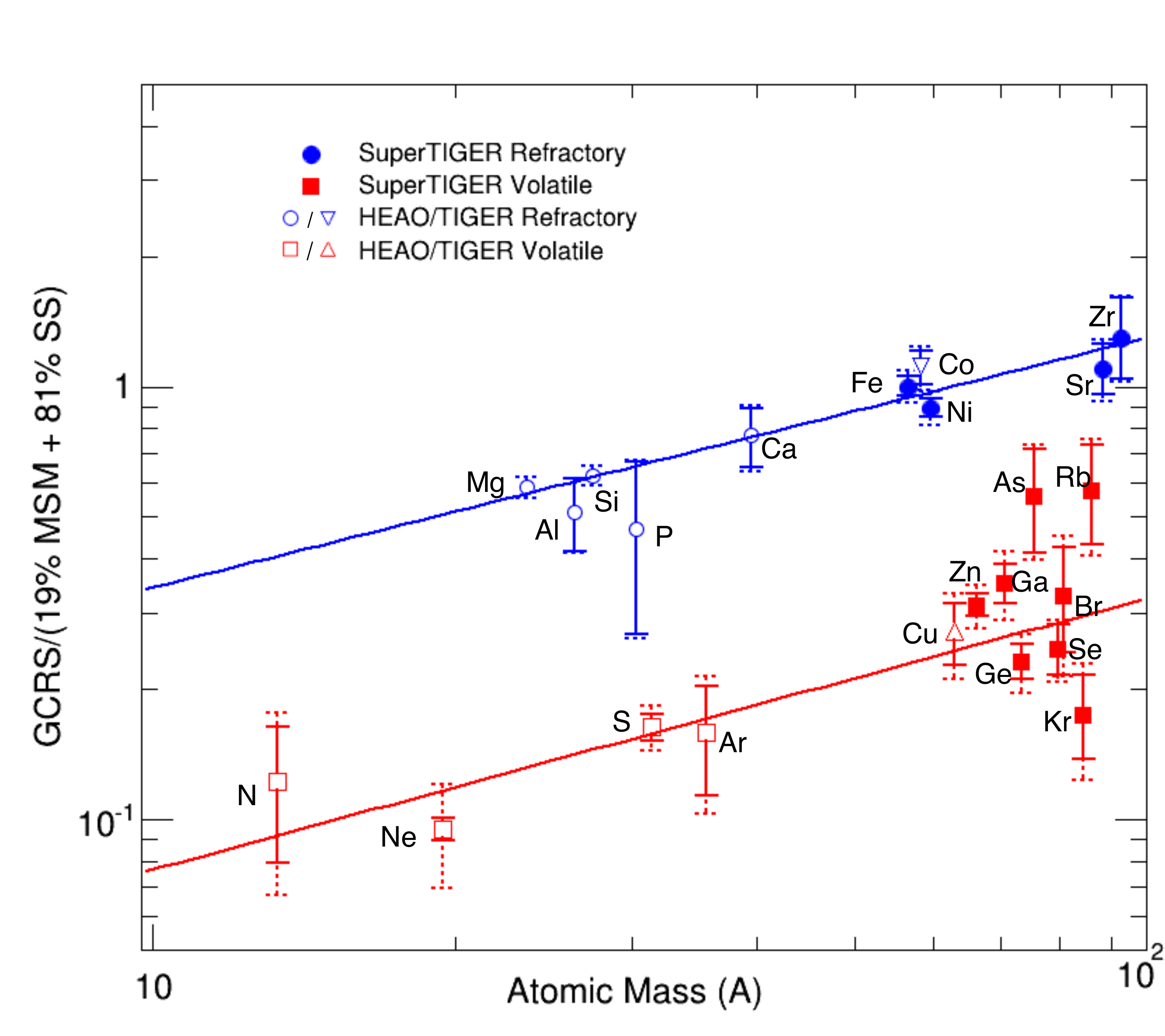}
\caption{This figure is the same as Figure \ref{GCRS_SS}, except the reference abundances to which GCRS abundances are compared are a mixture of 19\% MSM \citep{WoosleyHeger} and 81\% SS abundances \citep{Lodders2003}. Solid error bars show the uncertainty in the ratio due to uncertainty in the GCRS measurement; dashed error bars show the total uncertainty in the GCRS/source mixture ratio, including uncertainty in the SS abundances.}
\label{GCRS_MSO}
\end{figure}

\begin{figure}[t]
\centering  
\includegraphics[width=0.5\textwidth]{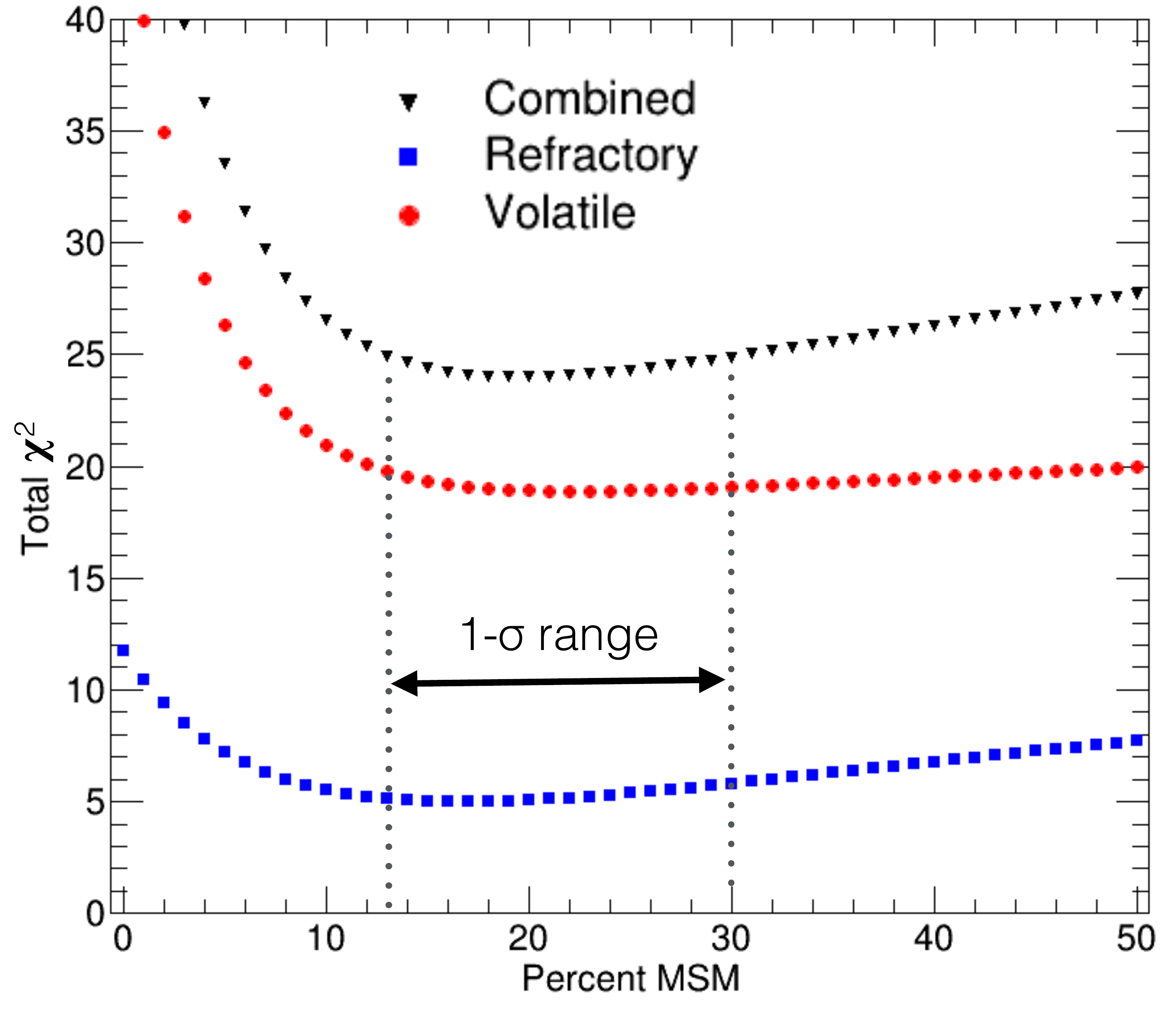}
\caption{Total $\chi^2$ value of fits as a function of the fraction of MSM from \citet{WoosleyHeger} included in the model source mixture. The minimum total $\chi^2$ for the combined refractory and volatile elements is at 19\% MSM by mass. Dotted lines show $\pm$1$\sigma$ errors in the best-fit mixture.}
\label{Chisq}
\end{figure}
Figure \ref{GCRS_MSO} is a modification of Figure \ref{GCRS_SS} where the GCRS abundances are compared to a mixture by mass of 81\% material with SS composition \citep{Lodders2003} and 19\% MSM averaged over an initial mass function (calculated by \citet{WoosleyHeger}), instead of pure SS material.
This figure shows a significant improvement in the organization of data compared with Figure \ref{GCRS_SS}, with a clear separation of the refractory and volatile elements, each with a similar mass dependence. The value of 19\% massive star material (MSM) was determined by comparing the GCRS abundances with source mixtures consisting of SS material with MSM mixed in 1\% increments from 0 to 100\%. For each source mixture, the refractory and volatile elements were each fit with a simple curve of the form $y=C_{0}A^{C_{1}}$, and the combined $\chi^2$ value for the mixture was calculated. The mixture with the minimum total $\chi^2$ was selected as the best fit mixture, with $\pm$1$\sigma$ uncertainty levels obtained by finding the mixtures with a $\chi^2$ value of $\chi_{min}^2$+1. The best-fit mixture was found to be 19$^{+11}_{-6}$\% MSM by mass, with the rest being normal interstellar medium (ISM) material with SS elemental abundances. Figure \ref{Chisq} shows the total $\chi^2$ for refractory elements, volatile elements, and the combined total $\chi^2$ as a function of the percentage of MSM in the source mixture. The vertical dashed lines show the $\pm$1$\sigma$ range in the percentage of MSM.

\begin{figure}[t]
\centering  
\includegraphics[width=0.5\textwidth]{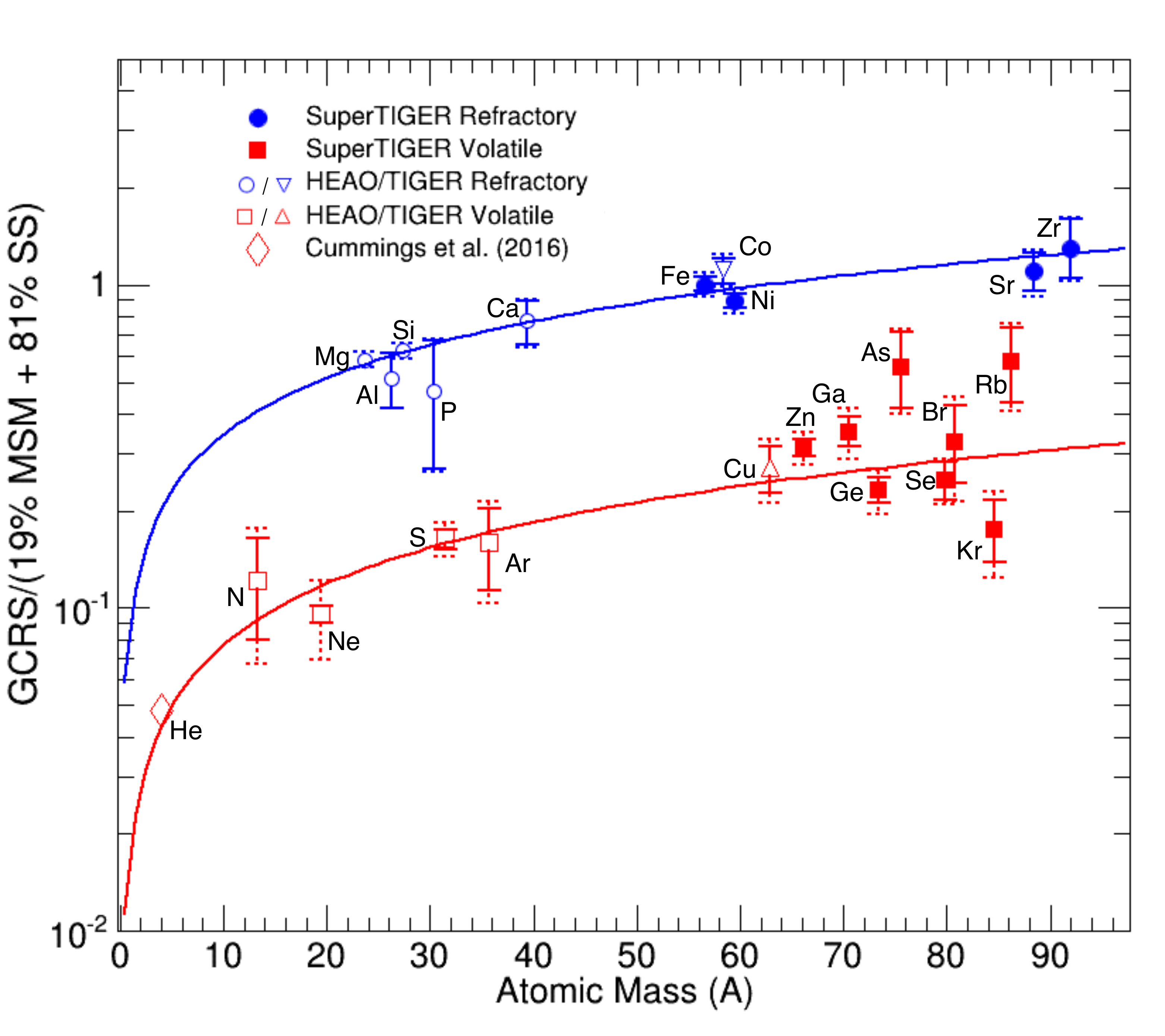}
\caption{This figure shows the same data and curves as Figure \ref{GCRS_MSO}, but with the addition of a $_2$He datum from \citet{Cummings}. The Atomic Mass axis is shown with a linear scale. $_2$He was not included in the fit, but still falls very near to the best-fit line. \citet{Cummings} do not report an uncertainty for $_2$He, so no error bars are shown for that point. For all other points, solid error bars show the uncertainty in the ratio due to uncertainty in the GCRS measurement; dashed error bars show the total uncertainty in the GCRS/source mixture ratio, including uncertainty in the SS abundances. }
\label{GCRS_MSO_He}
\end{figure}

Figure \ref{GCRS_MSO_He} uses the same data and curves as Figure \ref{GCRS_MSO}, but with the addition of a $_2$He datum from \citet{Cummings}. This point was not included in the fit, but the point still falls very near to the best-fit line. 
\begin{figure}[th!]
\centering  
\includegraphics[width=0.5\textwidth]{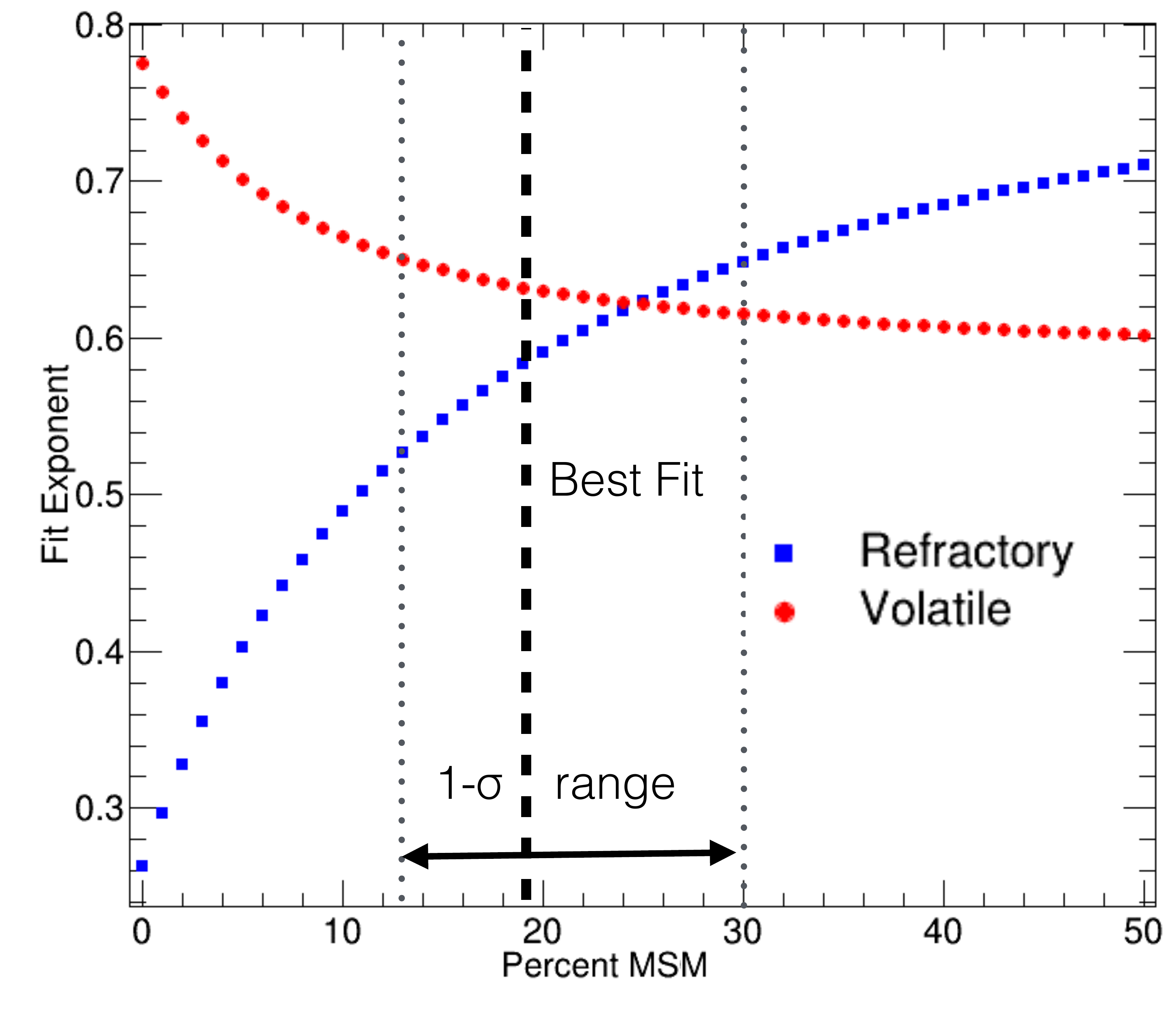}
\caption{Value of the slope C$_{1}$ for refractory and volatile elements as a function of the amount of MSM from \citet{WoosleyHeger} included in the model source mixture. For each mixture, both the refractory and volatile elements were fit with a curve of form $y=C_{0}A^{C_{1}}$. Dashed line shows the best-fit source mixture while dotted lines show the $\pm$1$\sigma$ uncertainty on that fit.}
\label{FitExponent}
\end{figure}

We use SS abundances from \citet{Lodders2003} instead of the more recent \citet{Lodders2009} because the massive star outflow and ejecta model of \citet{WoosleyHeger} used the \citet{Lodders2003} relative abundances as the input to their yield calculations. The more recent calculation of SN yields by \citet{Sukhbold} also uses \citet{Lodders2003}. We use the \citet{WoosleyHeger} results here instead of \citet{Sukhbold} because, as noted in their paper, the \citet{Sukhbold} SN yields are ``problematic" for elements in the ultra-heavy charge range. We note that yields have also been calculated by \citet{ChieffiLimongi}. Their yields for these UH elements differ from \citet{WoosleyHeger} and show strong dependence on the choice of mass cut, which determines how much material is ejected in the supernova explosion. A similar analysis should be done using their calculated yields. However, this is beyond the scope of the present paper.

This best fit had a reduced $\chi^2$ value of 1.26. The $C_1$ parameter of the best fit is 0.583 $\pm$ 0.072 for refractory elements and 0.632 $\pm$ 0.119 for volatile elements. Figure \ref{FitExponent} shows the fit $C_1$ values as a function of the percent of MSM in the source mixture. 

We interpret Figures \ref{GCRS_MSO} and \ref{GCRS_MSO_He} as strong evidence in support of the model of cosmic-ray origin in which the source material is a mix of MSM with normal ISM, and refractory elements are preferentially accelerated over volatile elements.  The contribution to the GCRS mixture from MSM indicates that OB associations are a significant source of GCR.
 
In addition, the recent detection of  $^{60}$Fe in cosmic rays ($^{60}$Fe is a radioactive primary cosmic ray with half-life 2.6 MYr that is primarily synthesized in core-collapse supernovae \citep{Travaglio,Seitenzahl,Sukhbold}) conclusively shows that recently synthesized material (within the last few million years) is accelerated to cosmic-ray energies \citep{Binns_60Fe}. The most natural place for this to occur is in OB associations.

These observations do not directly tell us where or how the normal ISM is injected and accelerated into cosmic rays. Type Ia supernovae often explode into normal ISM and it is estimated that 15\% of core-collapse supernovae occur outside of superbubbles \citep{Higdon2003,Higdon2005}. $\gamma$-ray observations show that these supernovae accelerate high-energy cosmic rays \citep{Dermer,Wang}. However, additional accelerators are required since an unreasonably large fraction of the supernova energy would be required to power cosmic rays if these were the only source of normal ISM acceleration. Since most supernovae are in OB associations, it appears that supernova shocks from stars in OB associations must also be accelerating cosmic rays from the normal ISM, perhaps from walls of superbubbles surrounding OB associations and residual ISM within the superbubble itself \citep{Higdon2003}. 
\section{Summary} \label{sec:Summary}
We have presented new SuperTIGER measurements of the elemental abundances of GCR from $_{26}$Fe to $_{40}$Zr.  Our results support a model of cosmic-ray origin in a source mixture of 19$^{+11}_{-6}$\% MSM and $\sim$81\% normal ISM material with solar system abundances. This indicates that a significant fraction of GCR acceleration occurs in OB associations. We also find a preferential acceleration of refractory elements over volatile elements by a factor of between $\sim$4 and $\sim$4.5, ordered by atomic mass (A). Both the refractory and volatile elements show a mass-dependent enhancement with similar slopes.

\acknowledgements
We gratefully acknowledge the excellent and highly professional work of the NASA Columbia Scientific Balloon Facility, the NASA Balloon Program Office, and the NSF Office of Polar Programs, who together made possible the record long-duration balloon flight of SuperTIGER. NASA supported this research under the ROSES 2007 APRA program under grants NNX09AC17G to Washington University in St. Louis and NNX09AC18G to the California Institute of Technology and the Jet Propulsion Laboratory, and APRA07-0146 to NASA/GSFC. We thank Nasser Barghouty for his help calculating interaction cross sections for the Galactic propagation. We are also grateful for support from the Peggy and Steve Fossett Foundation and the McDonnell Center for the Space Sciences at Washington University. We thank the referee who identified himself as Don Ellison for his helpful comments that improved this paper.




\listofchanges

\end{document}